\documentclass[12pt,a4paper]{scrartcl}
\usepackage[utf8]{inputenc}
\usepackage{booktabs}
\usepackage[table]{xcolor}
\usepackage{textcomp}
\usepackage{setspace}
\usepackage{threeparttable}
\usepackage{supertabular}
\usepackage{rotating}
\usepackage[longnamesfirst]{natbib}
\usepackage[]{csquotes}
\usepackage{amsmath}
\usepackage{amssymb}
\usepackage{eurosym}
\usepackage[margin=2.5cm]{geometry}
\usepackage{url}
\usepackage{authblk}
\usepackage{tikz}
\usepackage{pgfplots}
\pgfplotsset{compat=1.9}
\usepackage{multirow}
\newcommand{\tminus}{$_{t-1}$}

\usepackage{subcaption}

\begin{document}
\thispagestyle{empty}
\title{The Impact of Age on Nationality Bias: Evidence from Ski Jumping}

\author[1]{Sandra Schneemann}
\author[1]{Hendrik Scholten\thanks{%
corresponding author: hendrik.scholten@uni-bielefeld.de}}
\author[1]{Christian Deutscher}

\affil[1]{Bielefeld University, Department of Sport Science, Postfach 10 01 31, 33501 Bielefeld, Germany\thanks{%
For all authors: Declarations of interest: none}}
\renewcommand\Affilfont{\itshape\small}

\maketitle

\begin{abstract} 
This empirical research explores the impact of age on nationality bias. World Cup competition data suggest that judges of professional ski jumping competitions prefer jumpers of their own nationality and exhibit this preference by rewarding them with better marks. Furthermore, the current study reveals that this nationality bias is diminished among younger judges, in accordance with the reported lower levels of national discrimination among younger generations. Globalisation and its effect in reducing class-based thinking may explain this reduced bias in judgment of others.

\vspace{0.3cm} \noindent \textbf{JEL-Code}: 

\vspace{0.2cm} \noindent \textbf{Keywords}: subjective evaluation; in-group bias; own nationality bias; age

\end{abstract}

\newpage
\doublespacing

\section{Introduction}
Various professions must engage in evaluations of individual behaviours: Legal judges evaluate defendants’ actions, teachers grade students’ work, personnel managers evaluate applications and conduct performance reviews, and sports judges assess athletes’ competitive performance. All such evaluators are expected to assess others impartially and independent of personal characteristics, such as gender, race or nationality. Yet people’s assessments of others appear inherently biased, whether consciously or unconsciously, by the social identities of the parties involved. In(ter)group bias ``generally refers to the systematic tendency to evaluate one's own membership group (the ingroup) or its members more favourably than a nonmembership group (the outgroup) or its members'' \cite[p. 576]{hewstone2002}. In-group favouritism and discriminatory attitudes toward out-group members relate closely, leading to inequity in legal decisions, pay policies, and career chances — as well as championship titles. Understanding the existence and sources of in-group bias or out-group discrimination is critically important to be able to detect, reduce and mitigate the effects of, or even prevent such illegitimate judgments.

Biased assessments often are based on attributes such as race \citep{parsons2011,price2010}, gender \citep{blau2003understanding} or nationality \citep{emerson2009assessing}. These attributes can be observed easily and therefore serve as primary research objects for biased judgements. In general, people favour others of the same gender, race or nationality, as confirmed in substantial research that consistently illustrates evaluators’ preferences for people from the same country (or race or gender) over non-nationals (or other races or the opposite sex) \citep[see e.g.,][]{ahmed2013co,dee2005teacher,emerson2009assessing,zitzewitz2006}. 

In particular, extensive social science research investigates sources of out-group discrimination, revealing that both individual- and contextual-level factors affect discriminatory attitudes and ethnic prejudices. Age is one such variable; in general, younger people express less negative attitudes toward immigrants or people of ethnicities other than their own \citep[see e.g.,][]{orourke2006,quillian1995}.

Yet all investigations of nationality biases focus on direct comparisons of in- versus out-groups, without addressing how individual characteristics, such as age, might affect this bias. In an initial effort to close this gap, we empirically analyse in-group bias as it relates to the evaluator’s age. Thereby, we take advantage of the possibility to observe actual actions instead of only stated preferences. 

To determine the impact of age on nationality bias, we gather data from international ski jumping events. Jumpers are evaluated simultaneously by multiple judges from different countries, though judges often have the same nationality as the jumpers. Therefore, we can check statistically for the existence of an in-group bias among judges while controlling for the jumpers performance. Maybe more importantly, we can determine whether age has an impact on in-group bias. Our data, covering more than 14,000 jumps, strongly support the existence of an in-group bias; highlighting that this in-group bias increases with age but virtually disappears for judges who are younger than about 30-35 years of age.\par

We start by establishing the theoretical background for the existence of this in-group bias, derived from social identity theory. In Section 3, we review prior literature on in-group bias and sources of discriminatory attitudes. Following our empirical analysis of in-group bias and the moderating impact of age in a ski jumping contexts, we conclude with a discussion of the findings and some limitations that suggest directions for further research.

\section{Theoretical Background} \label{s:theory}
Discrimination refers to disadvantageous treatment of single persons or members of certain groups, as a manifestation of specific values, attitudes, prejudices, or unconscious associations. Social identity theory provides a fundamental theoretical basis explaining discrimination against (out-)groups, through its predictions about intergroup behaviour and social categorization. By analysing discrimination from a social-psychological perspective, \citet{tajfel1971social} help explain preferences for in-group members over out-group members. \citet[][p.24]{tajfel1982social} also offers a meaningful definition of social identity as “that part of the individuals’ self-concept which derives from their knowledge of their membership of a social group (or groups) together with the value and emotional significance of that membership.” 

Such group membership requires individuals to see themselves, and be seen by others, as belonging to a group. Citizens of the same nation thus can define themselves as members of the group defined by that nationality. In addition, each person compares his or her identified group and its values against those of other groups and develops a sense of prestige, which may be positive or negative, depending on whether the perceived discrepancy between the in- and the out-group evokes positive or negative values. In the search for self-satisfaction, people pursue membership in groups that they evaluate positively, so that they can enhance their self-image \citep{tajfel1979integrative}. 

\citet{tajfel1971social} demonstrate experimentally that out-group discrimination persists even for groups defined by trivial or random criteria. By requiring participants to distribute rewards to unknown in- or out-group members without any direct consequences for those participants, \cite{tajfel1971social} show that people prefer to assign greater rewards to the in-group, with a greater gap relative to the out-group, rather than maximize the rewards of both groups. That is, the mere fact of membership to a group (even a randomly assigned one) can trigger discrimination against non-members.

\section{Literature Review}

\subsection{Intergroup Bias: Biased Evaluations Based on Race and Nationality}

In- and out-group-biases have been documented repeatedly, in various forms and fields. \citet{ruffle2006}, for example, find members of an Israeli kibbutz, a collective community, are less corporative when paired with non-kibbutz members rather than anonymous kibbutz members. The bias persists not only when group members can choose voluntarily to join the group (e.g., kibbutz members) but also when the groups are defined by criteria that, for the most part, are inherent to the person at birth, such as gender, race or nationality.\footnote{\citet{bagues2017does} analyse competitions for professorship positions and determine that the presence of women in randomly assigned committees does not raise the success rates of women applying for such positions. But \citet{de2015gender} show that women have higher chances of being promoted by mixed-gender committees than by all-male committees. Many studies investigate the impact of gender on evaluations, with mixed results.}

Early evidence of a \textit{racial} bias by \citet{kraiger1985} comes from a meta-analysis of 74 studies, in which white raters consistently assign higher ratings to white ratees, as do black raters to black ratees. \citet{mount1997} confirm that race affects individual evaluations, such that in most cases, people favour others with whom they share the same evident race. In studying the effect of race on teachers’ perceptions of students, \citet{dee2005teacher} finds that teachers evaluate students’ behaviour more negatively when they do not share the same ethnical background. Dee notes that the region and students’ socioeconomic status affect this bias too. With data from U.S. trials, \citet{anwar2012} show that black defendants are convicted significantly more often than white defendants when the jury is composed of white members only. This negative effect of a white jury disappears if just one black person joins the jury pool. \citet{ahmed2013co} analyses data from a Swedish televised cooking show, in which contestants rate one another to determine a winner. Contestants with a Swedish origin award higher scores to contestants who share this nationality, relative to contestants with different ethnic backgrounds. In line with these results, \citet{price2010} note that NBA players are sanctioned (more personal fouls) when officiated by an opposite-race referee compared with an own-race refereeing crew.


Literature focused on \textit{nationality biases} confirms that evaluators prefer in-group members, or own-nationals, to out-group members, or non-nationals. \citet{link1998us} analyses whether the origins of authors and reviewers matter in rankings of submitted scientific articles. Distinguishing U.S. citizens from others, Link uncovers higher ratings for papers submitted by U.S. authors, especially if reviewers also come from the United States. In another academic setting, \citet{feld2016} consider evaluations of students by teachers, showing that students’ evaluations vary by nationality, though they attribute this effect to exophobia (favouritism of own-group) rather than to discrimination against students of a different nationality. \citet{glock2013does} find that German teachers are biased against Turkish students if the teachers see their stereotypes confirmed in some way. The nationalistic bias identified by \citet{acolin2016field} takes place in the French rental apartment market. When they sent applications via email, for which the names of the applicants suggested different ethnic backgrounds or nationalities, they determined that the fictional applicants with names suggesting a Northern African, Sub-Saharan African, or Turkish descent received 16 to 22 \% fewer responses than French applicants, signalling an own-nationality bias. In contrast, applications with Polish- or Portuguese/Spanish-sounding names did not suffer lower chances of receiving a response, which the authors explain by arguing that Polish and Portuguese/Spain immigrant groups have been integrated into French society. Therefore, the results suggest the primary effect of biased evaluations of out-group members.

Several other studies that address nationality biases consider sports contests. \citet{pope2015} use data from professional soccer to analyse refereeing activity during matches on the highest level of European club football (UEFA Champions League). They show that referees are about 10\% more likely to call a foul if the foul-suffering player is of the same nationality as the referee. These results hold in later rounds and among both inexperienced and experienced referees. Among dressage competitions, \citet{sandberg2016competing} identifies biased evaluations by judges for athletes from their own country, as well as for participants from countries that also are represented in the jury. However, \citet{hawson2010variability} do not find any favouritism in cases in which the judge and rider are from the same country. To the best of our knowledge, \citeauthor{zitzewitz2006}'s (\citeyear{zitzewitz2006}) study is the only one to analyse data from professional ski jumping. With observations from the 2001/2002 season (including the Olympics, World Cup events and World Championship contests), he finds an own-nationality bias that grows stronger in important competitions (e.g., Olympics, top-ranked contestants in a final round) and for team events. Yet Zitzewitz also finds that judges are biased against their nationalistic colleagues, leading to some compensation effects.

\subsection{(Racial) Attitudes and Age}
Extensive literature deals with racial or ethnic prejudices and their causes, defining ``Ethnic prejudices refer to generalised unfavourable opinions on one or more different ethnic outgroups'' \citep[][p. 1]{coenders2004}. Age, cohort and period effects drive changes in these public attitudes \citep{steeh1992}, though clear (statistical) separations according to age, cohort and period can be problematic or even ``not possible'' \citep[][p. 900]{glenn1976}. We consider how research has addressed both cohorts and ages. 

Broadly, older people tend to be more prejudiced than younger ones \citep{ceobanu2010, wilson1996}. Some authors attribute this trend to an age effect, but others emphasize cohort characteristics \citep{steeh1992,wilson1996}. For example, in a widely cited work, \citet{sears1983} argues that age or life stages can be important sources of attitude variations. Cognitive processing abilities, social support systems and attitude objects all are affected by age (or life stages), so younger and older people differ in their attitudes. \citet{sears1983} concludes that older people tend to preserve existing opinions and are more prone to reject new ideas, act in groups that support their existing attitudes and maintain rather “anachronistic” attitudes. Such tendencies might explain why older people often appear more prejudiced.

\citeauthor{hyman1956} (\citeyear{hyman1956}, \citeyear{hyman1964}), \citet{steeh1992} and \citet{wilson1996} instead focus on cohorts, rather than age effects. \citeauthor{hyman1956} (\citeyear{hyman1956}, \citeyear{hyman1964}) provide early empirical analyses of racial attitudes and postulate ``a model of positive change. In their view, a process of constant liberalization occurs as older, less tolerant generations are 'replaced' by new generations that have been socialised to be less hostile than their parents to racial integration'' \citep[][p. 341]{steeh1992}. In turn, researchers have related positive changes to assimilation and cited a ``long-term trend towards assimilation of ethnic minorities in modern societies'' \citep[][p. 406]{coenders1998}, resulting in less racial discrimination among younger cohorts. 

In their analysis of General Social Surveys and National Election Studies with respect to (changes in) racial attitudes, \citet{steeh1992} explicitly exclude ageing effects, arguing that existing evidence on age effects is not convincing. Rather, they propose that ``racial attitude trends are composed, therefore, only of cohort and period effects'' \citep[][p. 344]{steeh1992}, though their results indicate that neither cohort nor period effects significantly affect racial attitudes.

Finally, \citeauthor{wilson1996}'s (\citeyear{wilson1996}) analysis of cohort effects involves attitudes toward Blacks, Hispanics, Jews and Asians, as expressed in General Social Survey data. Unlike \citet{steeh1992} he finds cohort-specific differences, such that respondents born before versus after World War II differ significantly in their racial prejudices: Those born earlier stereotype minorities more adversely than those born in later years.

More recent studies of the determinants of racial prejudices and attitudes toward ethnic minorities repeatedly confirm that older respondents are more prone to discriminatory attitudes \citep[see e.g.,][]{coenders1998,gorodzeisky2009,orourke2006,quillian1995}. However, even recent studies raise some questions about whether age really relates to such opinions \citep[see e.g.,][]{mclaren2007,scheve2001}. Furthermore, most research does not specify whether the anticipated relationship reflects a cohort or an age effect, though a few explicitly assert that the relationship between age and attitudes may be due to both types of effects \citep{coenders1998}. Thus, the question of whether age or cohort characteristics induce racial prejudices remains unresolved, even though a general consensus indicates that older people tend to be more prejudiced than younger ones.

On the basis of this evidence, we predict that a potential nationality bias would be affected by evaluators’ age (specific socialisation): The older the judge, the more likely she or he prefers competitors of the same nationality. With an empirical analysis, we investigate this relationship using professional ski jumping data. First, we assess whether judges favour in-group nationals or discriminate against out-group non-nationals. Second, we analyse the relationship between judges’ ages and their exhibited nationality biases. Thus, instead of observing stated preferences our study builds on the investigation of actual actions .

\section{Empirical Analysis}
\subsection{Data}
To investigate the existence of in-group bias and the role of age, we rely on data from performance evaluations in professional ski jumping competitions. A jumper’s performance depends on two scores: distance and style points. Distance points are an objective measure of the length of the jump; style points reflect subjective assessments by judges. Depending on a jump’s distance, style points account for 40–50\% (mean = 46\%) of a jumper’s total score.\footnote{The longer a jump, the more distance points a jumper receives, which lowers the share of style points relative to the total score. The maximum number of style points a judge can award per jump is fixed at 20; distance points theoretically are not limited, though natural limits result from the construction of the jumping hills and the effects of gravity.} Thus, the style points fundamentally determine competitive outcomes.

Style points should reflect various aspects of a jump (e.g., landing, symmetry and steadiness of the body position during flight), awarded by five judges who watch each jump from the same spot next to the hill and rate it simultaneously. Each judge assigns exactly one \textit{style point} value to each jump. The best and worst scores out of the five get eliminated, so the jumper’s final style score represents the aggregation of the remaining three scores. Rough guidelines for scoring exist, yet judges have substantial room for discretion when evaluating jumpers’ performances. The highest rating an athlete can receive from a single judge is 20, and lower scores can be awarded at 0.5 increments.

Usually, the five judges come from five different countries — one judge from the country where the competition takes place and four from national ski jumping associations, selected by the international ski jumping federation in advance.\footnote{In one competition, two judges came from the same country, Norway (February 6, 2016, in Oslo).} The selection of judges is up to each national association, and any judges with valid certifications may be nominated. In order to qualify for a certification as a jumping judge candidates must have at least three years of practical experience as a national-level judge and must not exceed the age of 45. Regardless of the fact that having been a ski-jumper oneself is not obligatory for candidates, most of them do anyway. However, this does not imply only international careers as several of the judges in our sample reported national-level competition to be the highest level of competition they took part in.

Our data set includes all individual World Cup competitions for the 2015/16 and 2016/17 seasons, covering 41 competitions. Each competition usually consists of two rounds, in which (typically) 50 athletes start in the first round,\footnote{Usually, a qualification occurs prior to the competition, to determine the starting field of 50 athletes.} and then the top 30 qualify for the second round. Thus, we typically observe 80 jumps per competition, though the actual number can vary. For example, bad weather conditions might mandate the cancellation of the second round, or disqualifications might alter the number of jumpers allowed to start in the first round. Overall, the initial data set includes 3,203 different jumps, though we needed to exclude a few observations from our analysis (see the Appendix for a detailed criteria for these exclusions). We include only observations of judge-nationalities with at least two different judges,\footnote{If only one judge comes from a certain country, the estimated nationality effect may be due not to favouritism of own-nationals but rather individual-specific attitudes.} meaning that some jumps were rated by four instead of five judges. Finally, even though the best and worst ratings get excluded from the competitive score, we include them in our analysis, because these ratings may be particularly strong evidence of biased evaluations. Therefore, the data set for the analysis covers 3,025 jumps and 14,180 judge–jump observations; it includes 99 (82) athletes (judges), representing 16 (15) different countries.\footnote{The Appendix provides further details about the represented countries and observations.}

\subsection{Variables}

To assess judges’ subjective evaluations of individual jumps, our dependent variable is the number of \textit{style points} an individual judge awards to a specific jump. The dummy variable \textit{samenation} indicates whether the jumper and judge share a nationality. Then the main independent variables indicate judges’ \textit{ages} and the interaction of \textit{age} and \textit{samenation}. To analyse age-specific effects on in-/out-group bias, we measure \textit{age} as the judges’ ages at the time of the competition.

Several control variables also allow us to account for jump-, jumper- and judge-specific factors. Generally, longer jumps are better, and better jumps often earn more style points, so we must include jump distance in the analysis. It provides a good indicator of the jumper’s overall performance. Rather than the absolute length of the jump though, we use the number of \textit{distance points} in our analysis to control for overall performance, because ski jumping hills vary with respect to their size. Therefore, any given distance could constitute good or bad performance. \textit{Distance points} take hill size into account as it provides a better proxy of a jumper’s performance. We also include the squared term of \textit{distance points}, because exceptionally long jumps are hard to land accurately, due to the flatter landing areas, which may result in lower \textit{style points}.

Another parameter of jumpers’ performance involves wind conditions. Strong winds, for example, make it more difficult to maintain a good posture during the flight. Unfortunately, information on the actual wind direction and force per jump are not available. However, we have data on the \textit{wind points} each jumper receives for each jump. These points compensate jumpers for good or bad wind conditions at the time of their jump,\footnote{Being upwind positively affects distance; tailwinds typically lead to shorter jumps.} so they represent a (rough) measure of wind conditions.

To determine whether a contest takes place in the jumper’s home country, we include the dummy variable \textit{home (jumper)}. Jumping “at home” likely increases the number of \textit{style points} for two reasons. First, social pressure generated by (the noise of) the crowd may influence the judges’ perceptions \citep{nevill1996factors}. Second, jumpers generally are more familiar with domestic ski jumping hills, because they have more opportunities to practise on them. Thus, we anticipate that jumpers generally perform better in home-country competitions.

In accounting for jumper- and judge-specific effects, we run our estimations with jumper and/or judge fixed effects. With robustness checks, we also include a variable to capture a jumper’s average style points in all previous jumps during the considered time period (\textit{style points$_{t-1}$}), instead of using jumper fixed effects.\footnote{Including both fixed effects and a lagged dependent variable may lead to biased estimators \citep{angrist2008mostly}, so we only include one or the other (\textit{style points$_{t-1}$} or jumper fixed effects).}

Table \ref{descr} displays descriptive statistics for the key variables.\footnote{Table \ref{corr} in the Appendix shows the correlation matrix for the independent variables.} The dependent variable \textit{style points} ranges between 14.5 and 20,\footnote{Ratings below 14.5 are possible but excluded from our analysis (see the Appendix for reasons). In both seasons we study, fewer than 1\% of all ratings fall below this score. In addition, we identify only 7 observations of a maximum score of 20.} with a mean value of 17.56. The descriptive statistics suggest that judges rate jumpers higher if they come from the same country (17.687 vs. 17.543). A t-test shows that this difference is significant at the 0.01 level.

\begin{table}[htbp]
  \centering
  \footnotesize
  \begin{threeparttable}
    \begin{tabular}{llcccccc}
    \toprule
    \multicolumn{2}{l}{\textbf{Variable}} & \textbf{Obs} & \textbf{Level} & \textbf{Mean} & \textbf{Std. Dev.} & \textbf{Min} & \textbf{Max} \\
    \midrule
    \multicolumn{2}{l}{\textit{style points}} & 14180 & jump-judge & 17.557 & 0.822 & 14.5  & 20 \\
          & samenation & 1356  & jump-judge & 17.687$^*$ & 0.778 & 14.5  & 19.5 \\
          & other nation & 12824 & jump-judge & 17.543$^*$ & 0.826 & 14.5  & 20 \\
    \midrule
    \multicolumn{2}{l}{\textit{samenation}} & 14180 & jump-judge & 0.096 & 0.294 & 0     & 1 \\
    \multicolumn{2}{l}{\textit{age (judge)}} & 14180 & jump-judge & 48.666 & 8.882 & 22.174 & 64.063 \\
    \midrule
    \multicolumn{2}{l}{\textit{home (jumper)}} & 3025  & jump  & 0.121 & 0.326 & 0     & 1 \\
    \multicolumn{2}{l}{\textit{distance points}} & 3025  & jump  & 63.654 & 13.744 & 7.8   & 106.8 \\
    \multicolumn{2}{l}{\textit{distance points$^2$}} & 3025  & jump  & 4240.716 & 1720.019 & 60.84 & 11406.24 \\
    \multicolumn{2}{l}{\textit{wind points}} & 3025  & jump  & -0.774 & 9.006 & -28.8 & 24.6 \\
    \textit{style points$_{t-1}$} &       & 3025  & jump  & 17.489 & 0.528 & 15.200 & 18.929 \\
    \midrule
    \multicolumn{2}{l}{\textit{total style points$^{**}$}} & 3025  & jump  & 52.689 & 2.314 & 44    & 59 \\
    \multicolumn{2}{l}{\textit{total points$^{***}$}} & 3025  & jump  & 115.930 & 17.960 & 45.9  & 167.7 \\
    \bottomrule
    \end{tabular}%
    \begin{tablenotes} \scriptsize
    \item[*] T-test of \textit{style points} same vs. other nation: significant at 0.01 level.
    \item[**] \textit{total style points} refer to the sum of the three scores that remain after eliminating of the best and worst ratings. 
    \item[***] \textit{total points} = \textit{distance points} + \textit{total style points} + \textit{wind points} + \textit{gate points}.
    \end{tablenotes}
\end{threeparttable}
  \caption{Descriptive Statistics}
  \label{descr}%
\end{table}%

Overall, about 10\% of all jumps are rated by judges who share a nationality with the jumper. Jumps are heterogeneous in their distance points.\footnote{This measure refers to a hill-specific construction point (K-Point) that serves as a starting point for calculating the distance points. The K-Point position depends on hill size and slope, varying between 90 m (normal hill) and 130 m (large hill) in our data set. A jumper who jumps exactly at the K-Point receives 60 distance points. Depending on the hill size, a jumper receives between 1.8 and 2.0 points for each additional metre. If the jump is shorter than the K-Point, points are subtracted respectively from the starting 60 points.} Because \textit{wind points} compensate for wind conditions, the respective values can be negative (above-average wind conditions) or positive (below-average wind conditions). Even though only professional athletes compete, the jumpers’ skills also vary considerably, such that \textit{style points}$_{t-1}$ ranged between a minimum of 15.2 and a maximum of 18.9.

\begin{figure}[htbp]
\centering
\begin{minipage}{0.5\textwidth}
\centering
\scriptsize
\begin{tikzpicture}[baseline=(current axis.north)]
\begin{axis}[
width=1\textwidth,height=0.25\textheight,
ybar,
bar width=9pt,
ymin=0, ymax=26,
axis lines=middle,
ylabel={\textbf{Percentage}},
ytick={0, 5, ..., 25},
xtick={14, 14.5, ..., 20},
xmax=20.5, xmin=14,
ymajorgrids,
y label style={at={(ticklabel* cs:1.03)}, anchor=south, right=0mm, align=left},
]
\addplot[black!80,fill=gray]
table{
14.5	0.22
15	0.62
15.5	1.27
16	3.98
16.5	9.49
17	17.4
17.5	23.98
18	23.6
18.5	13.6
19	4.99
19.5	0.79
20	0.05
};
\draw[black] (axis cs:0,0) -- (axis cs:3.15,0);
\end{axis}
\end{tikzpicture}
\subcaption{\textit{style points}} \label{histoPoints}
\end{minipage}
\begin{minipage}{0.4\textwidth}
\scriptsize
\centering
\begin{tikzpicture}[baseline=(current axis.north)]
\centering
\begin{axis}[
width=1\textwidth,height=0.25\textheight,
ybar,
bar width=9pt,
ymin=0, ymax=31,
axis lines=middle,
ylabel={\textbf{Percentage}},
ytick={0, 5, ..., 30},
xtick={15, 20, ..., 65},
xmax=63, xmin=15,
ymajorgrids,
y label style={at={(ticklabel* cs:1.03)}, anchor=south, right=0mm, align=left},
]
\addplot[black!80,fill=gray]
table{
20	1.64
25	1.96
30	4.9
35	6.35
40	15
45	14.89
50	28.94
55	22.25
60	4.08
};
\draw[black] (axis cs:0,0) -- (axis cs:3.15,0);
\end{axis}
\end{tikzpicture}%
\subcaption{\textit{age}} \label{histoAge}
\end{minipage}
\caption{Histograms - \textit{style points} and \textit{age}}
\label{f:histo}
\end{figure}
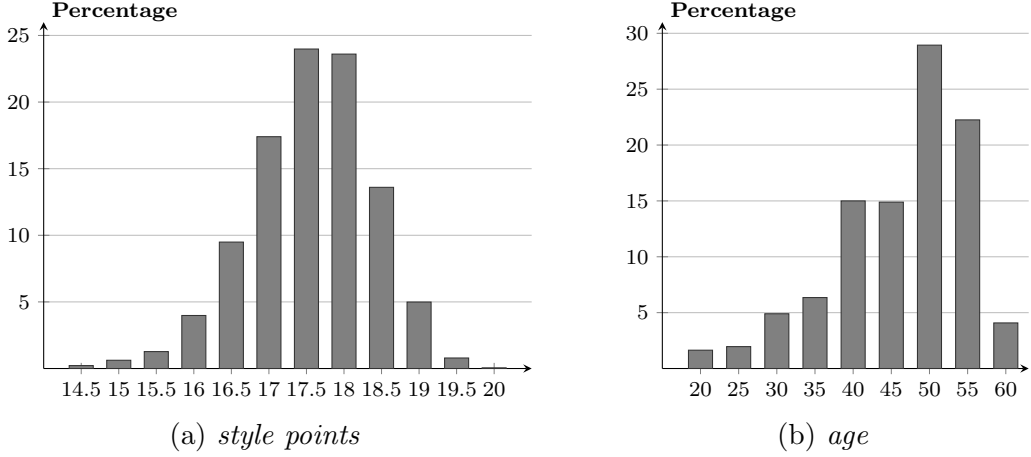

Figure \ref{f:histo} shows the distribution of the dependent variable \textit{style points} (Figure \ref{histoPoints}) and the independent variable \textit{age} (Figure \ref{histoAge}). Whereas \textit{style points} is rather normally distributed, age tends to be left skewed. Only about 15\% of observations come from judges younger than 40 years of age, and more than 80\% are from judges between the ages of 40 and 60 years. 

\subsection{Empirical Analysis of Nationality Bias}

We first focus on the existence of a nationality bias and therefore exclude the \textit{age (judge)} variable and its interaction term \textit{samenation * age}. Using both ordinary least squares and fixed effects regressions, we estimate the following equation:
\vspace{-0.1cm}
\begin{equation}
\small
\begin{split}
\textit{style points}_{ijt} = \beta_0 & +\beta_1~\textit{samenation}_{ijt} 
+ \beta_2~\textit{home (jumper)}_{it} + \beta_{3}~\textit{distance points}_{it} \\
				   & + \beta_{4}~\textit{distance points}_{it}^2+ \beta_{5}~\textit{wind points}_{it} + \beta_{6}~\textit{style points}_{it-1}  + u_{ijt}
\end{split}
\label{eq:reg}
\end{equation}

where \textit{style points}$_{ijt}$ represents the number of style points jumper \textit{i} receives from judge \textit{j} at time \textit{t}\footnote{This \textit{t} is a specific time on a specific date, because 30 jumpers perform two jumps on each day of competition.}.

The results of our analysis are in Table \ref{reg1}. We find strong support for an own-nationality bias, robust across different model specifications.\footnote{The Appendix contains several additional robustness checks. Table \ref{reg1b} presents results of Model 1 and 2 without using \textit{style points$_{t-1}$ (jumper)}. For Table \ref{reg2c} we did the same for Model 4.} On average and under otherwise identical conditions, judges award more points to jumpers from the same country than they do to others.\footnote{We further checked whether the strength of the own-nationality bias differs between the two rounds of the contests. We find that this is not the case. Thus, the favouritism of fellow-countrymen appears to be rather a subconscious and permanent effect than an explicit one which is only made use of when stakes are high.} The coefficient of \textit{samenation} is highly significant at the 1\% level in all three models, yet the coefficients of around 0.07 indicate that the extent of favouritism is rather small: It refers to only one-twelfth of the \textit{style points’} standard deviation. The impact of jumping at home \textit{(home (jumper))} tends to be approximately the same size as matched nationality with the judge. As we hypothesized, the distance a competitor jumps significantly affects the number of style points awarded: The better (longer) the jump, the higher the style score. Furthermore, we find the predicted reversed U-shaped relationship between \textit{distance} and \textit{style points}.

\begin{table}[htbp]
  \centering
  \footnotesize
  \begin{threeparttable}
    \begin{tabular}{lccc}
    \toprule
          & \textbf{Model 1} & \textbf{Model 2} & \textbf{Model 3} \\
    \midrule
    \textit{samenation} & 0.0668*** & 0.0764*** & 0.0775*** \\
          & (0.0122) & (0.0118) & (0.0115) \\
    \textit{home (jumper)} & 0.0783*** & 0.0605*** & 0.0784*** \\
          & (0.0224) & (0.0216) & (0.0212) \\
    \textit{distance points} & 0.0691*** & 0.0623*** & 0.0680*** \\
          & (0.0051) & (0.0053) & (0.0055) \\
    \textit{distance points$^2$} & -0.0002*** & -0.0002*** & -0.0002*** \\
          & (0.0000) & (0.0000) & (0.0000) \\
    \textit{wind points} & -0.0014* & -0.0017* & -0.0003 \\
          & (0.0009) & (0.0009) & (0.0009) \\
    \textit{style points$_{t-1}$ (jumper)} & 0.3486*** & 0.3208*** &  \\
          & (0.0206) & (0.0205) &  \\
    \midrule
    Constant & 8.0863*** &       &  \\
          & (0.3724) &       &  \\
    \midrule
    Jumper FE & no    & no    & yes \\
    Judge FE & no    & yes   & yes \\
    \midrule
    Obs.  & 14180 & 14180 & 14180 \\
    R2    & 0.606 & 0.640 & 0.674 \\
    adj. R2 & 0.606 & 0.638 & 0.670 \\
    \bottomrule
    \end{tabular}%
    \begin{tablenotes} \scriptsize
    \item Notes: The dependent variable is \textit{style points}. Robust standard errors are in parentheses.
    \item * p$<$0.1, ** p$<$0.05, *** p$<$0.01 (clustered on the jump level)
    \end{tablenotes}
\end{threeparttable}
  \caption{Regression Analysis: Own Nationality Bias}
  \label{reg1}%
\end{table}%

\subsection{Impact of Age on Nationality Bias}

Extending the existing literature, we now turn to the impact of age on national biases in judgments by including the variables \textit{age} and \textit{samenation * age} in the estimations. The results in Table \ref{reg2b} reveal that judges’ ages affect the extent of in-group favouritism:\footnote{We do not run estimations with jumper or judge fixed effects, because judges’ ages are relatively constant over the considered time period, so judge-specific and age effects tend to be collinear.} Older judges are more likely to award more style points to jumpers of the same nationality. In contrast, young judges even tend to award fewer points to jumpers of their own nationality. According to Model 4 (5), judges older than 36 (31) years discriminate in favour of jumpers from their home country, whereas those older than about 60 years award approximately 0.2 style points more to in-group competitors compared with out-group jumpers. These results resonate with previous research in other fields that show that older people from majority population express more negative attitudes toward immigrants and ethnic minorities \citep{gorodzeisky2009,orourke2006}. In international ski jumping events, the favouritism of fellow nationals is especially strong among older judges.


\begin{table}[h!]
  \centering
  \footnotesize
  \begin{threeparttable}
    \begin{tabular}{lcc}
    \toprule
          & \textbf{Model 4} & \textbf{Model 5} \\
    \midrule
    \textit{samenation} & -0.2817*** & -0.2071** \\
          & (0.0841) & (0.0812) \\
    \textit{samenation * age} & 0.0071*** & 0.0056*** \\f  
          & (0.0017) & (0.0016) \\
    \textit{age (judge)} & 0.0008* & 0.0012*** \\
          & (0.0004) & (0.0004) \\
    \midrule
    \textit{home (jumper)} & 0.0745*** & 0.0959*** \\
          & (0.0224) & (0.0229) \\
    \textit{distance points} & 0.0693*** & 0.0744*** \\
          & (0.0051) & (0.0052) \\
    \textit{distance points$^2$} & -0.0002*** & -0.0003*** \\
          & (0.0000) & (0.0000) \\
    \textit{wind points} & -0.0014 & -0.0008 \\
          & (0.0009) & (0.0008) \\
    \textit{style points$_{t-1}$ (jumper)} & 0.3486*** &  \\
          & (0.0206) &  \\
    \midrule
    Constant & 8.0407*** &  \\
          & (0.3725) &  \\
    \midrule
    Jumper FE & no    & yes \\
    Judge FE & no    & no \\
    \midrule
    Obs.  & 14180 & 14180 \\
    R2    & 0.607 & 0.638 \\
    adj. R2 & 0.606 & 0.635 \\
    \bottomrule
    \end{tabular}
    \begin{tablenotes} \scriptsize
    \item Notes: The dependent variable is \textit{style points}. Robust standard errors are in parentheses.
    \item * p$<$0.1, ** p$<$0.05, *** p$<$0.01 (clustered on the jump level)
    \end{tablenotes}
\end{threeparttable}
  \caption{Regression Analysis: Impact of Age on Nationality Bias}
  \label{reg2b}%
\end{table}%

The control variables (cf. \textit{wind points}) are highly significant in both model specifications, with the expected signs. Jumpers benefit from competition sites in their home countries, such that they receive significantly more style points in these competitions, potentially due to the jumpers’ familiarity with the hill or judges’ favouritism expressed to satisfy the “home crowd”. The results for \textit{distance points} and \textit{distance points}$^2$ also confirm the expected positive relation, with decreasing marginal effects. Overall, the R-square values range between 0.606 and 0.674, so the chosen variables explain a considerable portion of variance.

\section{Conclusion}

This study seeks two main research objectives, with the aid of recent data from professional ski jumping events. First, we test for and confirm the existence of an own-nationality bias, which in turn confirms the predictions based on social identity theory. Second, we provide the first test of whether nationally biased judgments are affected by judges’ ages. Previous studies indicate that attitudes tend to be more biased among older people or age cohorts; this study is the first to analyse if this effect holds when it comes to favouritism toward people from the same country in high-stakes judging situations. In support of our predictions, we find that the older a judge, the more athletes from the same country receive favourable ratings and the more athletes from other countries are discriminated against. The bias appears to be subconscious since data shown no strategic use of favoring jumpers in decisive jumps. The finding that in-group bias increases with age might signal a consequence of globalisation, such that self-concept definitions \citep{tajfel1982social} rely less on nationality among younger age cohorts. 

Professional ski jumping offers a useful and pertinent natural experiment, yet some limitations also exist in this study. First, the distribution of the \textit{age (judge)} variable is left skewed; only a few judges in our sample are younger than 40 years. However, the high number of observations from each judge strengthen the reliability of the results. Second, we have no data about individual-level factors (e.g., education) that also are likely to affect discriminatory attitudes. Third, our measure of wind conditions relies solely on the strength of tailwinds and headwinds; data about sidewinds are not available, though such conditions also could have effects. Fourth, career concerns might lead young judges to exhibit more conservative behaviours and judgements of jumpers from their same country, to avoid allegations of favouritism \citep{gibbons1992optimal}. More experience and increased job security might reduce this effect, parallel with increasing age.

This study presents initial insights into the impact of age on discriminatory judgements in a high stakes environment; other research gaps also need to be addressed in this context. For example, a data set that covers a longer period of time could reveal individual developments in judging patterns by single judges as they get older. In addition, it would be interesting to test for the effects of gender on the impact of age on nationally biased judgements. In our data set, nearly all the judges are men, and a gender variable thus would have measured individual judging patterns, not gender effects. If more women start to judge professional ski jumping, this field could provide a compelling research setting for analysing gender effects.


\clearpage
\bibliographystyle{apalike}
\bibliography{nationalitybias}

\clearpage

\section*{Appendix}

\subsection*{Data Cleaning and Exclusions} \label{exclusion}
\begin{enumerate} 
\item If a jump is very short, the jumper typically “abandons” the jump and does not worry about a good landing, resulting in very low ratings. Therefore, we exclude all jumps whose minimum score is below 14.5. \vspace{-0.3cm}
\item We exclude all jumps ending in a fall, because the number of style points depends crucially on the fall and does not reflect the quality of the previous flight phase. \vspace{-0.3cm}
\item If a jump’s minimum and maximum style point scores differ by more than 2 points, we exclude it; this situation suggests that one or more judges overlooked a major mistake (e.g., gripping into the snow on landing). \vspace{-0.3cm}
\item Some observations must be excluded because data about the judge’s age were missing. \vspace{-0.3cm} 
\item We only include observations of nationalities with at least two judges. If only one judge represents a country (as occurred for \textit{Slovakia}, \textit{Korea} and \textit{Kazakhstan}), the estimated nationality effect may be due not to favouritism of own-nationals but rather to individual-specific attitudes. 
\end{enumerate}

\clearpage

\subsection*{Countries of Jumpers} \label{a:jumper}
\begin{table}[htbp]
  \centering
  \footnotesize
    \begin{tabular}{rlcccc}
    \toprule
          & \multirow{2}{*}{\textbf{Nationality}} & \textbf{No. of} & \multirow{2}{*}{\textbf{Freq.}} & \multirow{2}{*}{\textbf{Percent}} & \multirow{2}{*}{\textbf{Cum.}} \\
          & & \textbf{Jumpers} & & & \\ \midrule
    1     & AUT   & 13    & 2,003  & 12.9  & 12.9 \\
    2     & BUL   & 1     & 50    & 0.32  & 13.22 \\
    3     & CAN   & 1     & 190   & 1.22  & 14.44 \\
    4     & CZE   & 7     & 1,199  & 7.72  & 22.17 \\
    5     & EST   & 1     & 36    & 0.23  & 22.4 \\
    6     & FIN   & 5     & 425   & 2.74  & 25.14 \\
    7     & FRA   & 2     & 409   & 2.63  & 27.77 \\
    8     & GER   & 11    & 2,250  & 14.49 & 42.26 \\
    9     & ITA   & 3     & 176   & 1.13  & 43.39 \\
    10    & JPN   & 10    & 1,405  & 9.05  & 52.44 \\
    11    & NOR   & 9     & 2,060 & 13.27 & 65.98 \\
    12    & POL   & 9     & 1,940 & 12.49 & 78.48 \\
    13    & RUS   & 6     & 680   & 4.38  & 82.86 \\
    14    & SLO   & 13    & 1,793 & 11.55 & 94.4 \\
    15    & SUI   & 5     & 630   & 4.06  & 98.46 \\
    16    & USA   & 3     & 239   & 1.54  & 100 \\ \midrule
          & \textbf{Total} & 99    & 14,180 & 100   &  \\
    \bottomrule
    \end{tabular}%
  \label{jumper}%
\end{table}%

\clearpage

\subsection*{Countries of Judges} \label{a:judges}
\begin{table}[htbp]
  \centering
  \footnotesize
    \begin{tabular}{rlcccc}
    \toprule
          & \multirow{2}{*}{\textbf{Nationality}} & \textbf{No. of} & \multirow{2}{*}{\textbf{Freq.}} & \multirow{2}{*}{\textbf{Percent}} & \multirow{2}{*}{\textbf{Cum.}} \\
          & & \textbf{Judges} & & & \\
    \midrule
    1     & AUT   & 10    & 1,538 & 10.85 & 10.85 \\
    2     & CAN   & 2     & 435   & 3.07  & 13.91 \\
    3     & CZE   & 3     & 459   & 3.24  & 17.15 \\
    4     & FIN   & 7     & 1,337 & 9.43  & 26.58 \\
    5     & FRA   & 2     & 375   & 2.64  & 29.22 \\
    6     & GER   & 18    & 2,224 & 15.68 & 44.91 \\
    7     & ITA   & 2     & 513   & 3.62  & 48.53 \\
    8     & JPN   & 3     & 818   & 5.77  & 54.29 \\
    9     & NOR   & 9     & 1,761 & 12.42 & 66.71 \\
    10    & POL   & 7     & 1,259 & 8.88  & 75.59 \\
    11    & ROU   & 2     & 387   & 2.73  & 78.32 \\
    12    & RUS   & 4     & 782   & 5.51  & 83.84 \\
    13    & SLO   & 7     & 1,151 & 8.12  & 91.95 \\
    14    & SUI   & 4     & 677   & 4.77  & 96.73 \\
    15    & USA   & 2     & 464   & 3.27  & 100 \\
    \midrule
          & \textbf{Total} & 82    & 14,180 & 100   &  \\
    \bottomrule
    \end{tabular}%
  \label{judges}%
\end{table}%

\clearpage

\subsection*{Correlation Matrix}

\begin{table}[htbp]
  \centering
  \footnotesize
  \begin{threeparttable}
    \begin{tabular}{l|cccccc}
    \toprule
          & \textit{samen.}  & \textit{age} & \textit{home} & \textit{distance p.} & \textit{wind p.} & \textit{style\tminus} \\
    \midrule
    \textit{samenation} & 1    &       &       &       &       &  \\
    \textit{age} (judge) & 0.0155  & 1     &       &       &       &  \\
    \textit{home} & 0.1389 &  0.0304 & 1     &       &       &  \\
    \textit{distance points} & 0.0209  & 0.0131 & 0.0071 & 1     &       &  \\
    \textit{wind points} & 0.0142 &  -0.0682 & 0.029 & -0.0603 & 1     &  \\
    \textit{style points\tminus} & 0.0411 &  -0.0001 & 0.0227 & 0.4425 & 0.0446 & 1 \\
    \bottomrule
    \end{tabular}%
\end{threeparttable}
  \caption{Correlation Matrix of Independent Variables}
  \label{corr}%
\end{table}%

\clearpage
\subsection*{Regression 1: Excluding $\mathit{style}$ $\mathit{points}$\tminus $\mathit{(jumper)}$}

Because \textit{style points}$_{t-1}$ correlates slightly with \textit{distance points} (see Table \ref{corr}) , the longer a jump, the better the jumper typically is, so this jumper likely has received higher average style points in prior competitions. Accordingly, we reran Models 1 and 2 from Table \ref{reg1} without \textit{style points}$_{t-1}$. The results are robust: \textit{samenation} again is significant at the 0.01 level, and its coefficient even higher than in Table \ref{reg1}. The impact of wind conditions seems small.

\begin{table}[htbp]
  \centering
  \footnotesize
  \begin{threeparttable}
    \begin{tabular}{lcc}
    \toprule
          & \textbf{Model 1b} & \textbf{Model 2b} \\
    \midrule
    \textit{samenation} & 0.0852*** & 0.0950*** \\
          & (0.0127) & (0.0123) \\
    \textit{home (jumper)} & 0.0859*** & 0.0635*** \\
          & (0.0237) & (0.0229) \\
    \textit{distance points} & 0.0712*** & 0.0640*** \\
          & (0.0053) & (0.0056) \\
    \textit{distance points$^2$} & -0.0002*** & -0.0001*** \\
          & (0.0000) & (0.0000) \\
    \textit{wind points} & 0.0001 & 0.0011 \\
          & (0.0009) & (0.0010) \\
    \textit{style points$_{t-1}$ (jumper)} &    --   &  --\\
          &       &  \\
    \midrule
    Constant & 13.9171*** & 14.0522*** \\
          & (0.1565) & (0.1660) \\
    \midrule
    Jumper FE & no    & no \\
    Judge FE & no    & yes \\
    \midrule
    Obs.  & 14180 & 14180 \\
    R2    & 0.566 & 0.607 \\
    adj. R2 & 0.566 & 0.605 \\
    \bottomrule
    \end{tabular}%
    \begin{tablenotes} \scriptsize
    \item Notes: The dependent variable is \textit{style points}. Robust standard errors are in parentheses.
    \item * p$<$0.1, ** p$<$0.05, *** p$<$0.01 (clustered on the jump level)
    \end{tablenotes}
\end{threeparttable}
  \caption{Regression Analysis: Own Nationality Bias, Excluding \textit{style points$_{t-1}$ (jumper)}}
  \label{reg1b}%
  \end{table}

\clearpage
\subsection*{Regression 2: Excluding $\mathit{style}$ $\mathit{points}$\tminus $\mathit{(jumper)}$}

As we did for Model 1 and Model 2 we also reran Model 4 from Table \ref{reg2b} without \textit{style points$_{t-1}$ (jumper)} due to the slight correlation with \textit{distance points}. Again, \textit{samenation} as well as the interaction term \textit{samenation * age} prove to be robust to the model modification as both are significant at the 0.01 level.

\begin{table}[htpb]
  \centering
  \footnotesize
  \begin{threeparttable}
    \begin{tabular}{lcc}
    \toprule
          & \textbf{Model 4b} \\
    \midrule
    \textit{samenation} & -0.2743***\\
          & (0.0872) & \\
    \textit{samenation * age} & 0.0073*** \\
          & (0.0017) \\
    \textit{age (judge)} & 0.0005 \\
          & (0.0005) \\
    \midrule
    \textit{home (jumper)} & 0.0824***\\
          & (0.0237) \\
    \textit{distance points} & 0.0713*** \\
          & (0.0053) \\
    \textit{distance points$^2$} & -0.0002*** \\
          & (0.0000)\\
    \textit{wind points} & -0.0001\\
          & (0.0009)\\
    \midrule
    Constant & 13.8894*** &  \\
          & (0.1578) &  \\
    \midrule
    Jumper FE & no \\
    Judge FE & no \\
    \midrule
    Obs.  & 14180 \\
    R2    & 0.567\\
    adj. R2 & 0.566\\
    \bottomrule
    \end{tabular}
    \begin{tablenotes} \scriptsize
    \item Notes: The dependent variable is \textit{style points}. Robust standard errors are in parentheses.
    \item * p$<$0.1, ** p$<$0.05, *** p$<$0.01
    \end{tablenotes}
\end{threeparttable}
  \caption{Regression Analysis: Impact of Age on Nationality Bias, Excluding \textit{style points$_{t-1}$ (jumper)}}
  \label{reg2c}%
\end{table}%

\end{document}